\documentclass{aa}
\usepackage{natbib}
\usepackage{graphicx}
\usepackage{txfonts}
\setcitestyle{aysep={}}

\begin{document}

\title{GRB 071227: an additional case of a \textit{disguised} short burst}

\author{
L. Caito\inst{1,2}
\and
L. Amati\inst{2,3}
\and
M.G. Bernardini\inst{4,1,2}
\and
C.L. Bianco\inst{1,2}
\and
G. De Barros\inst{1,2}
\and
L. Izzo\inst{1,2}
\and
B. Patricelli\inst{1,2}
\and
R. Ruffini\inst{1,2,5}
}

\institute{
Dipartimento di Fisica and ICRA, Universit\`a di Roma ``La Sapienza'', Piazzale Aldo Moro 5, I-00185 Roma, Italy.
\and
ICRANet, Piazzale della Repubblica 10, I-65122 Pescara, Italy. E-mails: letizia.caito@icra.it; amati@iasfbo.inaf.it; maria.bernardini@icra.it; bianco@icra.it; gustavo.debarros@icranet.org; luca.izzo@icra.it; barbara.patricelli@icranet.org; ruffini@icra.it
\and
Italian National Institute for Astrophysics (INAF) - IASF Bologna, via P. Gobetti 101, 40129 Bologna, Italy.
\and
Italian National Institute for Astrophysics (INAF) - Osservatorio Astronomico di Brera, via Emilio Bianchi 46, I-23807 Merate (LC), Italy.
\and
ICRANet, Universit\'e de Nice Sophia Antipolis, Grand Ch\^ateau, BP 2135, 28, avenue de Valrose, 06103 NICE CEDEX 2, France.
}

\titlerunning{}

\authorrunning{Caito et al.}

\date{}

\abstract{
Observations of gamma-ray bursts (GRBs) have shown an hybridization between the two classes of long and short bursts. In the context of the fireshell model, the GRB light curves are formed by two different components: the \textit{proper} GRB (P-GRB) and the extended afterglow. Their relative intensity is linked to the fireshell baryon loading $B$. The GRBs with P-GRB predominance are the short ones, the remainders are long. A new family of \textit{disguised} short bursts has been identified: long bursts with a protracted low instantaneous luminosity due to a low density CircumBurst Medium (CBM). In the $15$--$150$ keV energy band GRB 071227 exhibits a short duration (about $1.8$s) spike-like emission followed by a very soft extended tail up to one hundred seconds after the trigger. It is a faint ($E_{iso}=5.8\times10^{50})$ nearby GRB ($z=0.383$) that does not have an associated type Ib/c bright supernova (SN). For these reasons, GRB 071227 has been classified as a short burst not fulfilling the Amati relation holding for long burst.
}
{
We check the classification of GRB 071227 provided by the fireshell model. In particular, we test whether this burst is another example of a \textit{disguised} short burst, after GRB 970228 and GRB 060614, and, for this reason, whether it fulfills the Amati relation.
}
{
We simulate GRB 071227 light curves in the \emph{Swift} BAT $15$--$50$ keV bandpass and in the XRT ($0.3$--$10$ keV) energy band within the fireshell model. 
}
{
We perform simulations of the tail in the $15$--$50$ keV bandpass, as well as of the first part of the X-ray afterglow. This infers that: $E_{tot}^{e^\pm}= 5.04\times10^{51}$ erg, $B=2.0\times10^{-4}$, $E_{P-GRB}/E_{aft} \sim 0.25$, and $\langle n_{cbm} \rangle = 3.33$ particles/cm$^3$. These values are consistent with those of ``long duration'' GRBs. We interpret the observed energy of the first hard emission by identifying it with the P-GRB emission. The remaining long soft tail indeed fulfills the Amati relation.
}
{
Previously classified as a short burst, GRB 071227 on the basis of our analysis performed in the context of the fireshell scenario represents another example of a \textit{disguised} short burst, after GRB 970228 and GRB 060614. Further confirmation of this result is that the soft tail of GRB 071227 fulfills the Amati relation.
}


\maketitle

\section{Introduction}

The classification of short and long GRBs was introduced before the discovery of GRB afterglows \citep{costa} on the basis of their observed duration: $\Delta T < 2$ s for the short GRBs and $\Delta T > 2$ s for the long ones \citep{k93}. A possible connection between duration and spectral hardness was also proposed \citep{k93,t98}. In addition, ``long'' GRBs display an evident hard-to-soft transition and a time lag between the peak luminosities in different contiguous energy bands, the so-called ``spectral lag'', absent in the ``short'' ones \citep{n02}. Following the afterglow discovery and the spectral distribution detected by the \emph{BeppoSAX} satellite, another characteristic was introduced to discriminate between ``short'' and ``long'' GRBs. The afterglow was observed for the ``long'' GRBs only, which were assumed to be generated in the collapse of massive stars \citep[the ``collapsars'', see][]{wb06} and to occur in star-forming regions. The crucial point in that scenario was that ``long'' GRBs are intrinsically connected with supernovae (SN) events. ``Short'' GRBs were instead assumed to originate in coalescing neutron star binaries \citep[see e.g.,][and references therein]{ga09}. The first anomaly of this classification was already present in \emph{BATSE} data, but was identified only years later by \citet{nb06}. They noticed the existence of hybrid sources with an occasional, softer, prolonged emission lasting tenths of seconds in the gamma-ray energy band, following an initial spike-like emission comprising an otherwise short burst. The observations by \emph{Swift} further infringed the short/long dichotomy in many ways: some short duration GRBs were observed to be followed by an X-ray afterglow (first GRB 050509B observed by \emph{Swift}, see \citealp{ge05}, and, two months later, GRB 050709 observed by \emph{HETE-2}, see \citealp{villasenor}); some nearby long-duration GRBs were also observed to be not associated with SN (GRB 060614, \citealp{d06,f06,ga06}, and GRB 060505, \citealp{xu09}).

GRB 071227 also presents some intriguing anomalies. As for the ``Norris and Bonnel'' GRBs, its BAT light curve shows in the $15$--$150$ keV range a multi-peaked structure lasting $T_{90}=(1.8\pm0.4)$ s, followed by an extended but much softer emission up to $t_0+100$ s \citep{D09}. A fading X-ray ($0.3$--$10$ keV) and a faint optical afterglow have also been identified. The optical afterglow emission allowed to measure its redshift, $z=0.383$, and therefore its isotropic equivalent energy, $E_{iso}=5.8\times10^{50}$ erg in $20$--$1300$ keV \citep{D09}. The observed X-ray and optical afterglow is superimposed on the plane of the host galaxy, at $(15.0\pm2.2)$ kpc from its center.

On the basis of these characteristics, GRB 071227 has been classified as a short burst. This statement is supported by other main features: {\bf 1)} If we consider the first and apparently predominant short-duration episode, it does not fulfill the Amati relation between the isotropic equivalent radiated energy of the prompt emission $E_{iso}$ and the cosmological rest-frame $\nu F_{\nu}$ spectrum peak energy $E_{p,i}$ \citep{A02,A06b,a07,A09}. {\bf 2)} The spectral lag of the first spike-like emission in $25$--$50$ keV to $100$--$350$ keV bands is consistent with zero \citep{sa07}. {\bf 3)} Multiwavelength observations performed over many days have displayed that there is no association with a Ib/c hypernova, the type of SN generally observed with GRBs, even if it is a nearby burst and its isotropic energy is compatible with that of other GRBs associated with them \citep{D09}, although the upper limits are not deep enough to rule out a low-energetic core-collapse event. Nevertheless, the explosion of this burst in a star-forming region of a spiral galaxy, and its prolonged tail of emission, makes it most likely to be a long burst.

In this paper, we show that all these ambiguities and peculiarities can be explained in the framework of the fireshell model if we assume GRB 071227 to be a \textit{disguised} short burst, in which the first spike-like emission coincides with the P-GRB and the prolonged softer tail with the peak of the extended afterglow emitted in a low CircumBurst Medium (CBM) density region. We show, moreover, that this tail satisfies the Amati relation, and this is consistent with our interpretation.

\section{Brief summary of the \textit{fireshell} model}\label{model}

Within the fireshell model \citep{R02,R04,R05,R09,B05,Bia05}, all GRBs originate from an optically thick $e^\pm$ plasma of total energy $E_{tot}^{e^\pm}$ in the range $10^{49}$--$10^{54}$ ergs and a temperature $T$ in the range $1$--$4$ MeV. After an early expansion, the $e^\pm$-photon plasma reaches thermal equilibrium with the engulfed baryonic matter $M_B$ described by the dimensionless parameter $B=M_{B}c^{2}/E_{tot}^{e^\pm}$, which must be $B < 10^{-2}$ to allow the fireshell to expand further. As the optically thick fireshell composed of $e^\pm$-photon-baryon plasma self-accelerates to ultrarelativistic velocities, it finally reaches the transparency condition. A flash of radiation is then emitted. This represents the proper-GRB (P-GRB). The amount of energy radiated in the P-GRB is only a fraction of the initial energy $E_{tot}^{e^\pm}$. The remaining energy is stored in the kinetic energy of the optically thin baryonic and leptonic matter fireshell that, by inelastic collisions with the CBM, gives rise to a multiwavelength emission. This is the extended afterglow.

\begin{figure}
\centering
\includegraphics[width=\hsize]{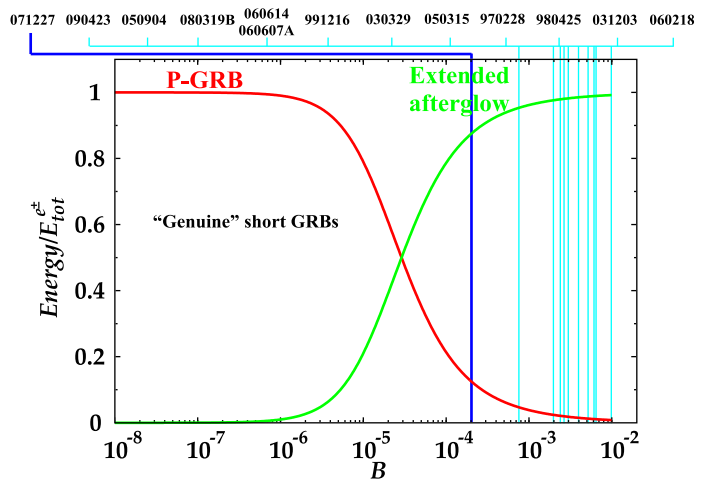}
\caption{The total isotropic energies emitted in the P-GRB (red line) and in the extended afterglow (green line), in units of the total energy of the plasma, as functions of the $B$ parameter. When $B \lesssim 10^{-5}$ the P-GRB becomes predominant over the extended afterglow, giving rise to a ``genuine'' short GRB. There are also marked in cyan the values of $B$ corresponding to some GRBs we analyzed, all belonging to the class of long GRBs, together with the GRB 071227 one (thick blue line).}
\label{f1}
\end{figure}

Within this model, the value of $B$ strongly affects the ratio of the energetics of the P-GRB to the kinetic energy of the baryonic and leptonic matter within the extended afterglow phase. It also affects the time separation between the corresponding peaks \citep{R09}. For baryon loading $B \lesssim 10^{-5}$, the P-GRB component is energetically dominant over the extended afterglow (see Fig. \ref{f1}). In the limit $B \rightarrow 0$, it gives rise to a ``genuine'' short GRB. Otherwise, when $10^{-4} \lesssim B \leq 10^{-2}$, the kinetic energy of the baryonic and leptonic matter, and consequently the extended afterglow emission, predominates with respect to the P-GRB \citep{R02,b07}.

The extended afterglow luminosity in the different energy bands is governed by two quantities associated with the environment: the CBM density profile, $n_{cbm}$, and the ratio of the effective emitting area $A_{eff}$ to the total area $A_{tot}$ of the expanding baryonic shell: ${\cal R}= A_{eff}/A_{tot}$. This second parameter takes into account the CBM filamentary structure \citep{R04} and the possible occurrence of fragmentation in the shell \citep{d07}. The emission from the baryonic matter shell is spherically symmetric. This allows us to assume, to first approximation, a modeling of thin spherical shells for the CBM distribution and consequently consider just their radial dependence \citep{R02}. The emission process is assumed to be thermal in the comoving frame of the shell \citep{R04}. The observed GRB non-thermal spectral shape is produced by the convolution of a very large number of thermal spectra with different temperatures and different Lorentz and Doppler factors. This convolution is performed over the surfaces of constant arrival time of the photons at the detector \citep[EQuiTemporal Surfaces, EQTSs;][]{B05,Bia05} encompassing the total observation time. The fireshell model does not address the plateau phase described by \citet{nu06}, which may not be related to the interaction of the single baryonic shell with the CBM \citep{ba10}.

The traditional fireball model addresses the details of the late afterglow phase by considering equations of motion expressed by power laws \citep[see e.g.][and references therein]{m06,ga09}. The fireshell model carefully considers the process of formation of the electron positron plasma, as well as the integration of the relativistic equations of motion from the early phases of the GRB to the late ``plateau'' phases. Special attention is given to the identification of the P-GRB. A highly nonlinear dependence on the above-mentioned parameters ($E_{tot}^{e^\pm}$, $B$, $n_{cbm}$, ${\cal R}$) has to be taken into account to obtain a single consistent solution within the fireshell model. The agreement with the observational data includes the luminosity in different energy bands as well as the spectra integrated over selected time intervals. The requirement of a single solution over all available data places a tight constraint on the parameter values.

\subsection{The new class of ``disguised'' short GRBs}

In the context of the fireshell model, we considered a new class of GRBs, pioneered by \citet{nb06}. This class is characterized by an occasional softer extended emission after an initial spike-like emission. The softer extended emission has a peak luminosity lower than the one of the initial spike-like emission. As shown in the prototypical case of GRB 970228 \citep{b07} and then in GRB 060614 \citep{c09}, we can identify the initial spike-like emission with the P-GRB and the softer extended emission with the peak of the extended afterglow. That the time-integrated extended afterglow luminosity is much larger than the P-GRB one is crucial. This unquestionably identifies GRB 970228 and GRB 060614 as canonical GRBs with $B > 10^{-4}$. The consistent application of the fireshell model allows us to infer the CBM filamentary structure and average density, which, in that specific case, is $n_{cbm} \sim 10^{-3}$ particles/cm$^3$ \citep{b07}. This low CBM density value explains the peculiarity of the low extended afterglow peak luminosity and its more protracted time evolution. These features are not intrinsic to the progenitor, but depend uniquely on the peculiarly low value of the CBM density. This led us to expand the traditional classification of GRBs to three classes: ``genuine'' short GRBs, ``fake'' or ``disguised'' short GRBs, and the remaining ``long duration'' ones.

A CBM density $n_{cbm} \sim 10^{-3}$ particles/cm$^3$ is typical of a galactic halo environment, and GRB 970228 was indeed found to be in the halo of its host galaxy \citep{Sahu97,VP97}. We therefore proposed that the progenitors of this new class of ``disguised'' short GRBs are merging binary systems, formed by neutron stars and/or white dwarfs in all possible combinations, which spiraled out from their birth place into the halo \citep[see][]{b07,c09,k06}. This hypothesis can also be supported by other observations. Assuming that the soft tail peak luminosity is directly related to the CBM density, GRBs displaying a prolonged soft tail should have a systematically smaller offset from the center of their host galaxy. Some observational evidence was found in this sense \citep{ta08}. However, the present sample of observations does not enable us to derive any firm conclusion that short GRBs with extended emission have smaller physical offsets than those without extended emission \citep{fa10,b10}.

\section{The interpretation of GRB 071227 light curves}\label{fit}

\begin{figure}
\centering
\includegraphics[width=\hsize]{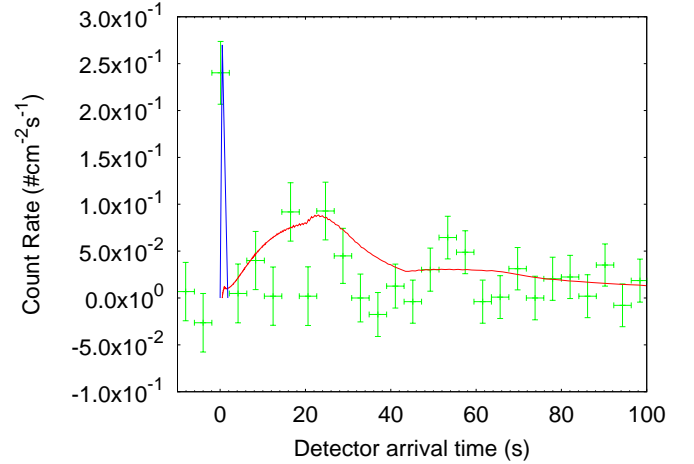}
\caption{The BAT $15$--$50$ keV light curve (green points) compared with the corresponding theoretical extended afterglow light curve (red line). The P-GRB is qualitatively sketched by the blue line.}
\label{f2}
\end{figure}

\begin{figure}
\centering
\includegraphics[width=\hsize]{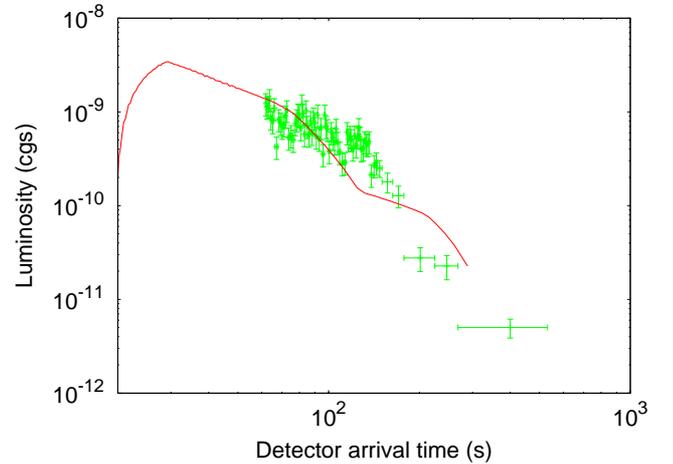}
\caption{The XRT $0.3$--$10$ keV light curve (green points) compared with the corresponding theoretical extended afterglow light curve we obtain (red line). The X-ray data corresponding to the first $60$ seconds are not available since XRT starts to observe only $60$ seconds after the BAT trigger. We stop our analysis at $\gamma \sim 5$ when our relativistic dynamical model can no longer be applied}
\label{f3}
\end{figure}

We examine the possibility that GRB 071227 can also be classified as a ``disguised'' burst. We analyzed the observed light curves of this burst in the $15$--$50$ keV bandpass, corresponding to the lowest band of the $\gamma$-ray emission, detected by the BAT instrument on the \emph{Swift} satellite, and in the $0.3$--$10$ keV energy band, corresponding to the X-ray component from the XRT instrument. To model the CBM structure, we assume that $n_{cbm}$ is a function of only the radial coordinate, $n_{cbm}=n_{cbm}(r)$ (radial approximation). The CBM is arranged in spherical shells of width $\sim 10^{15}-10^{16}$ cm arranged in such a way that the corresponding modulation of the emitted flux closely resembles the observed shape. We assumed that the first short spike-like emission represents the P-GRB and the $\gamma$-ray tail is the peak of the extended afterglow. We therefore began the simulation in such a way that the extended afterglow light curve begins in coincidence with the peak of the P-GRB (about $1$s), as shown in Fig. \ref{f2}. To reproduce the observational data and the energetics observed for the P-GRB emission ($E_{iso}\sim1.0\times10^{51}$ erg), we require the initial conditions $E_{tot}^{e^\pm}= 5.04\times10^{51}$ erg and $B=2.0\times10^{-4}$. In Figs. \ref{f2} and \ref{f3} we plot the comparison of the GRB 071227 BAT and XRT data with the theoretical extended afterglow light curves. We obtained a good result in the prompt emission for the $15$--$50$ keV bandpass (see Fig. \ref{f2}), while, for $0.3$--$10$ keV, we only succeeded in reproducing the first decaying part of the XRT light curve (see Fig. \ref{f3}). We assumed this to correspond to the possible onset of the ``plateau'' phase of the extended afterglow \citep{nu06}.

\begin{figure}
\centering
\includegraphics[width=\hsize]{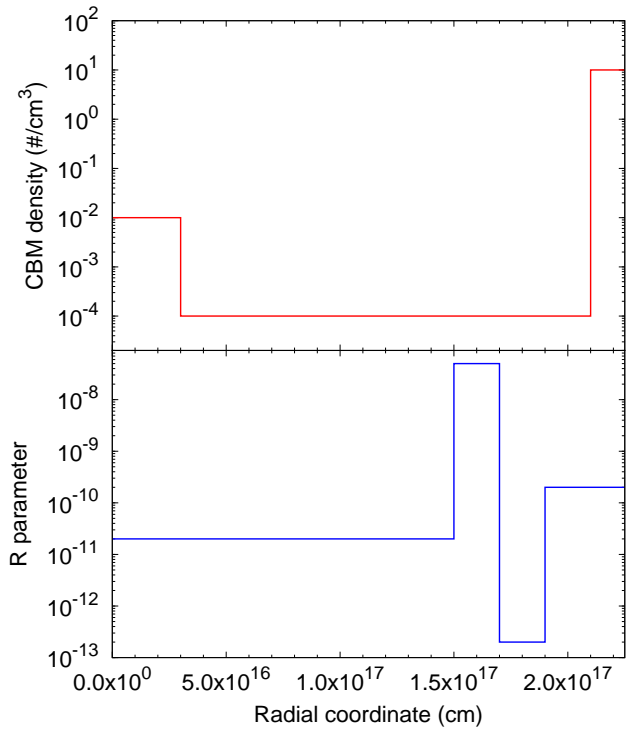}
\caption{The CBM particle number density $n_{cbm}$ (upper panel, red line) and the $\mathcal{R}$ parameter (lower panel, blue line) as functions of the radial coordinate}
\label{f4}
\end{figure}

From our simulation, the amount of energy stored in the P-GRB is found to be about $20$\% of the total energetics of the explosion. Hence, this burst cannot be a short burst within the fireshell scenario. The baryon loading obtained ($B=2.0\times10^{-4}$) remains in the range of long duration GRBs. This is a very critical value, because it is very close to the crossing point of the plot of the energetics of GRBs as a function of $B$ (see Fig. \ref{f1}). This is the lowest baryon loading that we have ever found in our analysis within the fireshell scenario. From our analysis, we found a peculiar result for the average CBM density. We obtained a density of $n_{cbm}=1.0\times10^{-2}$ particles/cm$^3$ at the beginning of the process, later decreasing to $n_{cbm}=1.0\times10^{-4}$ particles/cm$^3$. This low average density, inferred from the analysis, is responsible for the strong deflation of the $\gamma$-ray tail. However, at the radius of about $2.0\times10^{17}$ cm, the density becomes higher and reaches the value of $n_{cbm}=10$ particles/cm$^3$ (the complete profiles of $n_{cbm}$ and $\mathcal{R}$ as functions of the radial coordinate are reported in Fig. \ref{f4}). This is compatible with the observations. The observed X-ray and optical afterglow of GRB 071227 is indeed superimposed on the plane of the host galaxy, at $(15.0\pm2.2)$ kpc from its center \citep{D09}. An interesting possibility observed by D. Arnett (private communication) is that this very low density ``cavity'' could be formed in the coalescing phase of a binary formed by a neutron star and a white dwarf. Accurate studies on compact object mergers have shown the distribution of merger locations for different host galaxies \citep{ba06,b10}. In starburst galaxies, most of the mergers are expected to be found within hosts, while in elliptical galaxies a substantial fraction of mergers take place outside hosts. Spiral galaxies, hosting both young and old stellar populations, represent the intermediate case between the preceding two. This result is therefore compatible with our hypothesis about the binary nature of the progenitor of GRB 071227. Although they did not consider the case of binary systems formed by a neutron star and a white dwarf, the progenitor of GRB 071227 would fit into the tight binary scenario described by \citet{ba06}.

These results clearly imply that GRB 071227 is another example of a disguised burst.

\section{GRB 071227 within the Amati relation}\label{Amati}

One of the most effective tools for discriminating between ``short'' and ``long'' bursts, and possibly clarifying the interpretation of these different classes of events, is the Amati relation \citep{A02,A06b,a07,A09}.

This empirical spectrum-energy correlation states that the isotropic-equivalent radiated energy of the prompt emission $E_{iso}$ is correlated with the cosmological rest-frame $\nu F_{\nu}$ spectrum peak energy $E_{p,i}$: $E_{p,i}\propto (E_{iso})^{a}$, with $a \approx 0.5$ \citep{A02}. The existence of the Amati relation, discovered when analyzing the \emph{BeppoSAX} long duration bursts, has been confirmed by studying a sample of GRBs observed by \emph{Swift}, intense or soft, with available measurements of redshift and spectral parameters, and also by \emph{HETE-2} and \emph{Konus/WIND}. When the ``afterglow revolution'' allowed the redshift estimation of also some short GRBs, it was found that these bursts are inconsistent with the Amati relation, holding instead for long GRBs \citep{A06b,a10,aa09}. The most recent updating of the correlation (95 GRBs with the data available up to April 2009) also includes two high-energetic events detected by \textit{Fermi} (GRB 090323 and GRB 080916C), which are fully consistent with the $E_{p,i}$--$E_{iso}$ relation \citep{A09}. We also note that the short event GRB 090510, observed by the same satellite, is an outlier of the relation, as expected for short bursts. This dichotomy finds a natural explanation within the fireshell model. As recalled in Sect. \ref{model}, within this theoretical framework the prompt emission of long GRBs is dominated by the peak of the extended afterglow, while the one of the short GRBs is dominated by the P-GRB. Only the extended afterglow emission follows the Amati relation \citep[see][for details]{Guida}. Therefore, all GRBs in which the P-GRB provides a negligible contribution to the prompt emission (namely the long ones, where the P-GRB is at most a small precursor) fulfill the Amati relation, while all GRBs in which the extended afterglow provides a negligible contribution to the prompt emission (namely the short ones) do not.

\begin{figure}
\centering
\includegraphics[width=\hsize]{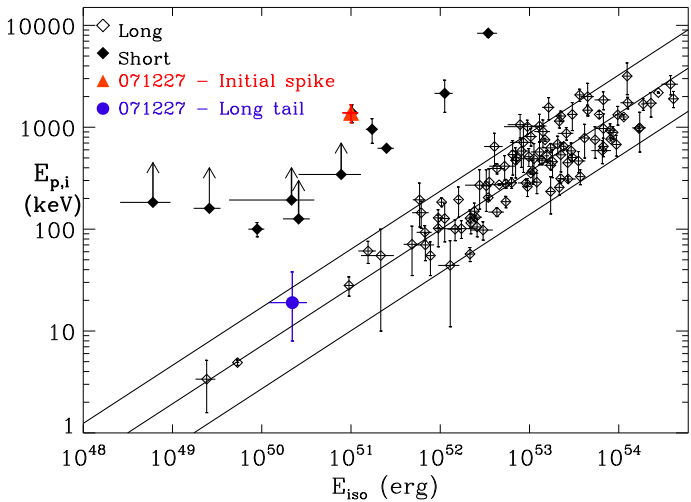}
\caption{Location of the initial short spike and soft long tail of GRB 071227 in the $E_{p,i}$ -- $E_{iso}$ plane. The data points of long GRBs are from \citet{a08,A09}, the data points and limits of short GRBs are from \citet{A06b,A09,p08}. The continuous lines show the best-fit power law and the 2$\sigma$ confidence region of the correlation, as determined by \citet{a08}.}
\label{f5}
\end{figure}

Apart from this general feature, there are peculiar cases. One of these is GRB 050724, a short burst followed by a long, softer tail in the $\gamma$-ray energy band. While the first short emission is inconsistent with the Amati relation, the soft tail is again consistent with it. Another one, the intriguing case of GRB 060614, a \textit{disguised} short burst within the fireshell scenario \citep{c09}, shows similar behavior: the first hard episode does not fulfill the Amati relation, while the whole event is fully consistent with it \citep{a07}. This is again fully consistent with the predictions of the fireshell model for the disguised GRB class. Since the first short and hard episode is the P-GRB and the prolonged softer tail is the peak of the extended afterglow, the Amati relation must be fulfilled only when the softer tail is considered alone, or considered together with the first episode if this episode provides a negligible contribution \citep[see][for details]{GraziaIC4}. Thus, to discriminate its nature, we studied the position of GRB 071227 in the $E_{p,i}$-$E_{iso}$ plane. At the observed redshift, we assumed a ``flat $\Lambda$-CDM model'' with $H_{0}=70 Km/s/Mpc$ and $\Omega_\Lambda=0.73$. For the first, hard spike, lasting about $1.8$s, using \emph{Konus/WIND} data \citep{ga07}, and integrating between $1$ and $10\, 000$ keV, we found that $E_{p,i}=(1384 \pm 277)$ keV and $E_{iso}=(1.0\pm0.2)\times10^{51}$ erg. As shown in Fig. \ref{f5}, this is inconsistent with the $E_{p,i}$-$E_{iso}$ relation, and instead occupies the short-populated region of the plane. For the long tail, lasting about $100$s, we used a Band model with $\alpha=-1.5$ and $\beta=-3$, which are typical of soft events. These values are compatible with the low quality statistics of this event. We found that $E_{p,i}=20_{-11}^{+19}$ keV and $E_{iso}=(2.2 \pm 0.1)\times10^{51}$ erg. With these values, the tail of emission is fully consistent with the Amati relation, as for any long GRB (see Fig. \ref{f5}). This clearly supports our hypothesis about the nature of GRB 071227.

\section{Conclusions}\label{conclusions}

GRB 071227 has been classified in the current literature as a short GRB not fulfilling the Amati relation. By analyzing it using the fireshell model, we have identified the first spike-like emission with the P-GRB and consequently the prolonged softer tail with the peak of the extended afterglow. We have shown that this tail indeed fulfills the Amati relation. We have shown that the Amati relation is a characteristic of the extended afterglow phase of GRBs and does not occur during the P-GRB emission. As shown in Fig. \ref{f1}, the relative energy of the P-GRB with respect to the extended afterglow is a strong function of the baryon loading. For $B\to 10^{-2}$, the energetic relevance of the P-GRB decreases and its contribution to the total GRB energetics can be neglected with respect to the extended afterglow: in this limit, the Amati relation always applies. In the present case, we have $B = 2.0\times 10^{-4}$ and the exclusion of the P-GRB from the total energetics of the GRB is indeed essential to fulfill the Amati relation. This is similar to the case of GRB 050724. With the considerations given above, it is appropriate to consider GRB 071227 to be another ``Norris and Bonnell'' kind of burst. It is quite similar to the bright GRB 060614, which we have previously analyzed \citep{c09}, although underluminous. The value of $B = 2.0\times 10^{-4}$ that we obtained appears to be the smallest of all GRBs we examined. It is particularly interesting that the CBM distribution is given by $\langle n_{cbm} \rangle < 10^{-2}$ particles/cm$^3$ to a radius of the order of $\sim 2.1\times 10^{17}$ cm and then monotonically rises to a value of $\langle n_{cbm} \rangle = 10$ particles/cm$^3$. While this last value is expected in the region where the GRB occurred, at $(15.0\pm2.2)$ kpc from the center of its host galaxy, the existence of a very low density ``cavity'' appears to be of great interest. On the other side, as for GRB 060614, the energetics of GRB 071227 is compatible with the progenitor being the merging of a binary system of a neutron-star and a white-dwarf. This binary system, during the long-lasting merging process, may have swept the CBM around by means of the pulsar radiation emission (D. Arnett, private communication). This remains under investigation and may be an additional factor affecting the analogous low density observed in GRB 060614. In that burst, the progenitor was also consistent with a binary system formed by a white dwarf and a neutron star.

\acknowledgements

We thank the Italian \emph{Swift} Team (supported by ASI grant I/011/07/0 and partly by the MIUR grant PRIN 2007TNYZXL) for the reduced \emph{Swift} data and Dr. Cristiano Guidorzi for his very important support in the data analysis. We thank as well an anonymous referee for her/his very important suggestions.

\end{document}